\documentclass[12pt]{l4dc2022}

\usepackage{amsfonts}
\usepackage{mathrsfs}
\usepackage{xspace}
\usepackage{xparse}

\usepackage{enumitem}
\usepackage{mathtools}
\usepackage{wrapfig}
\usepackage[font={small,it}]{caption}

\usepackage{algorithm}
\usepackage{algpseudocode}
\usepackage{setspace}
\usepackage{multicol}

\usepackage{booktabs}
\usepackage{multirow}

\usepackage{epstopdf}
\usepackage{lipsum}
\usepackage{color, soul}

\usepackage{pgfplots}

\pgfplotsset{compat=1.17}
\usetikzlibrary{tikzmark, calc}
\usetikzlibrary{arrows.meta}

\pgfmathdeclarefunction{gaussian_kernel}{3}{%
  \pgfmathparse{exp(-(abs(#1-#2)^2)/(2*#3^2))}%
}

\tikzset{ne/.style={circle, thick, inner sep=0pt, minimum size=3.0em}}
\tikzset{ns/.style={ne, draw}}
\tikzset{le/.style={draw, thick}}
\tikzset{ls/.style={le, -Latex}}

\tikzset{lnicebrace/.style={decorate, thick, line cap=round, decoration={brace, amplitude=5pt, raise=5pt}, yshift=0pt}}
\tikzset{rnicebrace/.style={decorate, thick, line cap=round, decoration={brace, mirror, amplitude=5pt, raise=5pt}, yshift=0pt}}

\tikzset{actualfn/.style={dot diameter=0.5pt, dot spacing=2pt, dots, black}}

\tikzset{sampledata/.style={only marks, color=black, mark=*, mark options={scale=0.5, fill=black}}}

\makeatletter
\tikzset{
    dot diameter/.store in=\dot@diameter,
    dot diameter=3pt,
    dot spacing/.store in=\dot@spacing,
    dot spacing=10pt,
    dots/.style={
        line width=\dot@diameter,
        line cap=round,
        dash pattern=on 0pt off \dot@spacing
    }
}
\makeatother

\newtheorem{defn}{Definition}[section]

\newtheorem{assumption}{Assumption}

\newcommand{\R}{\mathbb{R}}

\newcommand{\E}{\mathbb{E}}

\title[Chance Constrained Control using Kernel Distribution Embeddings]{Data-Driven Chance Constrained Control \\ using Kernel Distribution Embeddings}

\usepackage{times}

\author{%
 \Name{Adam J. Thorpe}
 \thanks{These authors contributed equally to this work.}
 \Email{ajthor@unm.edu}\\
 \addr Department of Electrical and Computer Engineering, University of New Mexico
 \AND
 \Name{Thomas Lew}
 \footnotemark[1]
 \Email{thomas.lew@stanford.edu}\\
 \addr Department of Aeronautics and Astronautics, Stanford University%
 \AND
 \Name{Meeko M. K. Oishi} \Email{oishi@unm.edu}\\
 \addr Department of Electrical and Computer Engineering, University of New Mexico
 \AND
 \Name{Marco Pavone} \Email{pavone@stanford.edu}\\
 \addr Department of Aeronautics and Astronautics, Stanford University%
}

\begin{document}

\maketitle

\vspace{-6mm}

\begin{abstract}
We present a data-driven algorithm for efficiently computing stochastic control policies for general joint chance constrained optimal control problems. Our approach leverages the theory of kernel distribution embeddings, which allows representing expectation operators as inner products in a reproducing kernel Hilbert space. This framework enables approximately reformulating the original problem using a dataset of observed trajectories from the system without imposing prior assumptions on the parameterization of the system dynamics or the structure of the uncertainty. By optimizing over a finite subset of stochastic open-loop control trajectories, we relax the original problem to a linear program over the control parameters that can be efficiently solved using standard convex optimization techniques. We demonstrate our proposed approach in simulation on a system with nonlinear non-Markovian dynamics navigating in a cluttered environment.
\end{abstract}


\section{Introduction}

The deployment of reliable autonomous systems requires control algorithms that are robust to model misspecifications and to external disturbances. 
To enable safety-critical applications, these control algorithms should also explicitly enforce constraints. 
For instance, an autonomous car should always respect speed limits and avoid pedestrians at all times while accounting for uncertain road conditions and external disturbances. 
The presence of these two sources of aleatoric and epistemic uncertainty presents a significant challenge for traditional stochastic optimal control techniques, which typically rely upon an accurate model of the system and calibrated uncertainty quantification. 
Critically, the accuracy of the model and of its associated uncertainty estimates may deteriorate over time as the system is deployed in new environments. 
This motivates the use of data-driven techniques that leverage collected measurements of the system to design efficient adaptive control laws. 
However, existing data-driven techniques tend to be complex to implement (e.g., neural network controllers), 
may require system-specific assumptions restricting possible applications (e.g., dynamics that are linear in the uncertain parameters), 
or may not explicitly account for constraints that are crucial for the reliable deployment of these intelligent autonomous systems.

\begin{figure}[t]
	    \centering
\includegraphics[width=0.6\linewidth,trim=0 0 0 0, clip]{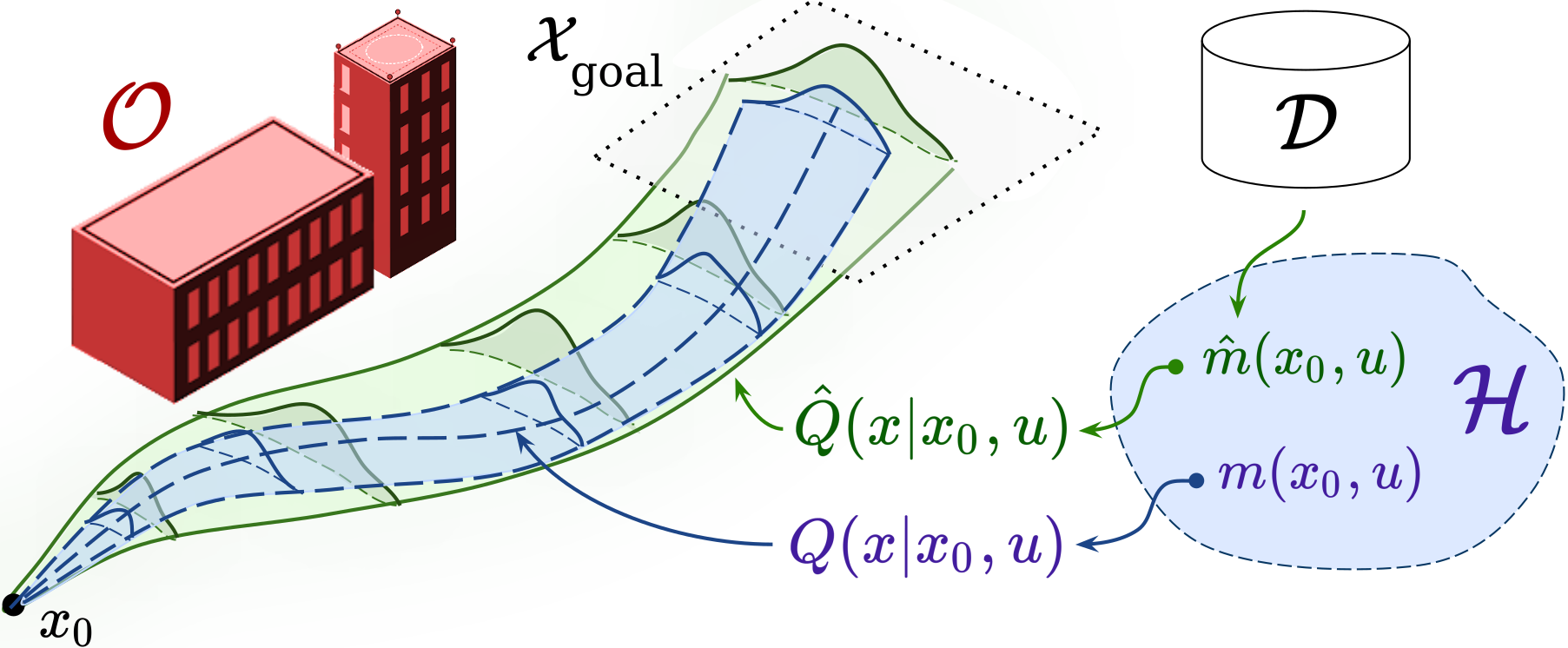}
	    \vspace{-2mm}
	    \caption{We propose a data-driven algorithm to efficiently compute control trajectories $u$ for uncertain dynamical systems. The key consists of representing the state trajectory distribution \textcolor{blue}{$Q(\cdot|x_0,u)$} by a conditional distribution embedding \textcolor{blue}{$m(x_0,u)$} from a reproducing kernel Hilbert space \textcolor{blue}{$\mathcal{H}$}. This approach allows approximating the true embedding \textcolor{blue}{$m(x_0,u)$} using a dataset $\mathcal{D}$ of observed trajectories of the uncertain system (see Assumption \ref{assum:dataset}). 
	    The resulting embedding estimate \textcolor[rgb]{0,0.4,0}{$\hat{m}(x_0,u)$} defines an approximation \textcolor[rgb]{0,0.4,0}{$\hat{Q}(\cdot|x_0,u)$} of \textcolor{blue}{$Q(\cdot|x_0,u)$} that is used to efficiently compute a stochastic control policy to drive the system to a goal region $\mathcal{X}_{\textnormal{\textrm{goal}}}$ while avoiding all unsafe sets \textcolor[rgb]{0.7,0,0}{$\mathcal{O}$} (e.g., representing obstacles) at all times.}
	    \vspace{-5mm}
	    \label{fig:overview}
	\end{figure}

\hspace{-4mm}\textbf{Contributions:} we present a data-driven control algorithm to efficiently compute stochastic control inputs for general joint chance constrained control problems given observed transitions from the system.
The key consists of leveraging the theory of \textit{kernel distribution embeddings}, which allows representing expectation operators as inner products in a reproducing kernel Hilbert space (RKHS). 
By applying this theory to the stochastic kernel that characterizes the uncertain system dynamics (Figure \ref{fig:overview}), we derive a tractable relaxation of the original joint chance constrained problem that is a linear program over the control parameters. 
Our approach allows computing randomized open-loop control strategies 
for non-Markovian nonlinear dynamical systems subject to nonconvex constraints. 

\textbf{Outline:} we discuss related work in Section \ref{sec:related_work} and our joint chance constrained problem formulation in Section \ref{section: problem formulation}. 
Section \ref{sec:stochastic_embeddings} describes how stochastic kernel embeddings are used to relax the original intractable problem using a dataset of trajectories as a linear program over the stochastic open-loop control parameters. 
We validate our approach in Section \ref{section: numerical results} and conclude in Section \ref{section: conclusion}.

\section{Related work}\label{sec:related_work}

A wide range of model-based stochastic control techniques have been developed to efficiently control uncertain systems. By leveraging a model of the system,
these approaches explicitly account for uncertainty while enforcing constraints.  
For instance,
stochastic model predictive controllers \citep{Mesbah2016} explicitly enforce constraints along the state trajectory,  
dynamic programming approaches \citep{OnoPavoneEtAl2015} use a Lagrangian relaxation and augment the control objective with a penalty on constraint violation, and control barrier functions \citep{Clark2019} provide a condition that
guarantees closed-loop forward invariance.
Guaranteeing constraint satisfaction with high probability typically involves considering a chance constrained problem formulation.   
The most common formulation enforces \textit{pointwise} chance constraints that ensure the independent satisfaction of each constraint at each time step with high probability \citep{CastilloRAL2020,
LewBonalli2020,
hewing2018cautious,
Polymenakos2020,
Khojasteh_L4DC20,
JasourRSS2021}. 
In contrast, joint chance constraints guarantee trajectory-wise constraints satisfaction with high probability \citep{blackmore2011chance,
FreyRSS2020,
SchmerlingPavone2017,
koller2018,
LewEtAl2021} which is of particular interest whenever all constraints should be satisfied at all times jointly. 
For instance, a drone transporting a package should always avoid obstacles and reach its destination 
with high probability over the distribution of possible payloads.
Unfortunately, tackling
joint chance constrained formulations is particularly challenging due to the need to consider the full distribution of the state trajectory. 
Common approaches decompose such constraints using Boole's inequality \citep{blackmore2011chance,Masahiro2016} which can be conservative \citep{SchmerlingPavone2017},
or perform robust trajectory optimization given confidence sets for the model parameters \citep{koller2018,LewEtAl2021} which requires assuming bounded external disturbances and relying on robust uncertainty propagation techniques that can be conservative.  

The performance of model-based controllers depends on prior domain knowledge,  
which may become inaccurate 
over time as the
system is deployed in new environments.  
For this reason, data-driven control techniques have been proposed to leverage measurements of the system to efficiently characterize its properties and adapt to changes in its operating conditions. 
Popular data-driven approaches leverage Gaussian processes models 
\citep{deisenroth2015,
Ostafew2016robust,
Khojasteh_L4DC20,
williams2006gaussian,
berkenkamp2017safe,
koller2018,
LewEtAl2021} which make Gaussian-distributed predictions for the state transitions of the system, 
linearly-parameterized dynamics \citep{Coulson2018DataEnabledPC,
Berberich2021}, 
Koopman operators \citep{Abraham2019}, random Fourier features \citep{BoffiRFF2021}, and neural networks \citep{Chua2018}. 
A key algorithmic feature of these approaches is to directly model single-step transitions of the system. 
While intuitive and sufficient from a statistical viewpoint, this approach 
makes uncertainty propagation challenging, requiring particle-based approaches to represent the distribution of the trajectory \citep{blackmore2010,JansonSchmerlingEtAl2015b,Chua2018}. 

In this work, we take a drastically different approach and leverage the theory of \textit{distributional kernel embeddings} to characterize the joint distribution of the state trajectory from data. This approach allows tackling general nonlinear, non-Markovian dynamical systems, representing joint chance constraints as a linear operation in an RKHS, and optimizing over open-loop stochastic control policies by solving a linear program over the control parameters. 
Kernel distribution embeddings have been thoroughly studied in the recent years \citep{song2009hilbert, song2007hilbert, grunewalder2012conditional, park2020measure}, but are not yet popular within the control community. 
Controller synthesis applications 
have been explored in \citep{thorpe2021stochastic}, but the authors do not consider constraints and only consider one-step transition kernels. 
In contrast, we tackle a joint chance constrained problem formulation that explicitly accounts for constraints and represent the distribution of the entire state trajectory. As a byproduct, this approach allows handling non-Markovian dynamical systems that are resistant to traditional one-step modeling techniques. 

\vspace{-2mm}

\section{Problem Formulation}
\label{section: problem formulation}

First, we describe our notation. 
We denote a conjunction (logical \scalebox{0.95}{\textrm{AND}}) by $\wedge$ and a disjunction (logical \scalebox{0.95}{\textrm{OR}}) by $\vee$. 
$\mathbb{R}$ and $\mathbb{N}$ denote the sets of real and natural numbers, respectively.  
Given a space $E$ and $N\in\mathbb{N}$, we denote the Cartesian product $E^N\triangleq E\times\dots\times E$ ($N$ times). 
For a subset $A \subset E$, the map $\boldsymbol{1}_{A} : E \to \lbrace 0, 1 \rbrace$ denotes the indicator function of $A$, which satisfies $\boldsymbol{1}_{A}(x) = 1$ if $x \in A$ and $\boldsymbol{1}_{A}(x) = 0$ if $x \notin A$. 
We denote the Borel $\sigma$-algebra on a topological space $E$ by $\mathscr{B}(E)$. 

\begin{defn}[Stochastic Kernel]\label{def:stochastic_kernel}
    \label{defn: stochastic kernel}
    Let $(E, \mathcal{E})$ and $(F, \mathcal{F})$ be measurable spaces. 
    A stochastic kernel from $E$ to $F$ is a map $\kappa : \mathcal{F} \times E \to [0, 1]$ such that $x \mapsto \kappa(A \mid x)$ is $\mathcal{E}$-measurable for all $A \in \mathcal{F}$ and $B \mapsto \kappa(B \mid x)$ is a probability measure on $(F, \mathcal{F})$ for all $x \in E$.
\end{defn}
In this work, we leverage stochastic kernels to describe the distribution of the state trajectory of a stochastic dynamical system, as a function of the initial state and of the selected control inputs.

\subsection{Chance Constrained Optimization}\label{sec:cc_problem}

Let $(\Omega, \mathcal{G}, \mathbb{P})$ be a probability space and $f:\mathbb{R}^n\times\mathbb{R}^m\times\mathbb{R}^p\times\mathbb{R}^q\rightarrow\R^n$ be a continuous function. 
Given an initial state $x_0\in\mathbb{R}^n$, our goal is to safely control a system with stochastic dynamics
\begin{equation}
    \label{eqn: system dynamics}
    x_{t+1} = f(x_t, u_t, w_t, \theta), \quad t\in\mathbb{N},
\end{equation}
where $x_t \in \mathcal{X} \subseteq \mathbb{R}^{n}$ denotes the state of the system at time $t$, 
$u_{t} \in \mathcal{U} \subset \mathbb{R}^{m}$ is the control input, 
$w=(w_t)_{t\in\mathbb{N}}$ is a stochastic process characterizing external disturbances,
and $\theta$ is a random variable that describes uncertain model parameters  (e.g., a drone transporting an uncertain payload of mass $\theta$ that is sampled at the beginning of the control task and held constant over time, see Section \ref{section: numerical results}).

This formulation is particularly challenging 
due to the fact that the system described in \eqref{eqn: system dynamics} is non-Markovian. Specifically, at any time $t\in\mathbb{N}$, the stochastic state trajectory $x=(x_t)_{t\in\mathbb{N}}$ satisfies
\begin{equation}\label{eq:dynamics:full}
    x_t = 
    f_t(x_0,u,w,\theta)
    \triangleq 
    f(\boldsymbol\cdot, u_{t-1}, w_{t-1}, \theta)
    \circ
    \dots
    \circ
    f(x_0, u_0, w_0, \theta),
\end{equation}
where $u=(u_s)_{s=0}^{t-1}$ denotes the control trajectory. 
Since the parameters $\theta$ are uncertain and the disturbances $w_t$ are not necessarily independent, the increments $x_{t+1}-x_t$ are not independent, i.e. the state trajectory $x$ is not a Markov process. Intuitively, the parameters are randomized only once and their uncertainty is propagated along the entire state trajectory, see \citep{dynsSDE2021}.

We consider the problem of minimizing the sum of two (possibly non-convex) state and input cost functions $\ell^{x}:\mathcal{X}^{N} \to \mathbb{R}$ and $\ell^{u}:\mathcal{U}^{N} \to \mathbb{R}$. For instance, $\ell^{x}$ may be chosen as a quadratic cost that penalizes tracking error and $\ell^{u}$ may penalize control effort. 
In addition to this control objective, the system should reach a desired compact goal region  $\mathcal{X}_{\textrm{goal}} \subset \mathcal{X}$ at a specified time $N\in\mathbb{N}$ while avoiding all unsafe sets of states $\mathcal{O}_{t} \subset \mathcal{X}$ (e.g., representing potentially nonconvex obstacles). 
Due to the stochasticity of the system, guaranteeing strict constraint satisfaction with probability one may be infeasible. Instead, given a tolerable failure probability threshold $\delta \in (0, 1)$, we require that all constraints are jointly satisfied  with probability at least $1-\delta$, a requirement often referred to as a joint chance constraint. 

In general, the optimal solution of such chance constrained control problems are \textit{stochastic} control policies \citep{Altman1999,Masahiro2016}.
As such, we formulate a stochastic optimal control problem over the set $\Delta$ of all open-loop stochastic policies 
$\pi: \mathscr{B}(\mathcal{U}^N) \times \mathcal{X} \to [0, 1]$, where
$\pi$  is a stochastic kernel from $\mathcal{X}$ to $\mathcal{U}^N$. 
Optimizing over closed-loop stochastic policies is challenging; we leave such extensions for future work. 
The resulting control problem is expressed as follows: 
\begin{subequations}
\label{eqn: chance constrained optimization problem}
\begin{align}
    \min_{\pi\in\Delta} \quad & \mathbb{E}[\ell^{x}(x) + \ell^{u}(u)] 
    \label{eq:problem:cost}
    \\
    \textnormal{s.t.} \quad & x_{t+1} = f(x_t, u_t, w_t, \theta),
    \quad 
    u\sim \pi(\cdot \mid x_0),
    \quad 
    t = 0, 1, \ldots, N-1, 
    \label{eq:problem:dynamics}
    \\
    \label{eqn:problem:joint_chance_constraint}
    & \mathbb{P} \biggl( \biggl( \bigwedge_{t=1}^{N-1} x_{t} \not\in \mathcal{O}_{t} \biggr) \land (x_{N} \in \mathcal{X}_{\textrm{goal}}) \biggr) \geq 1 - \delta,
\end{align}
\end{subequations}
where the state and control trajectories are denoted as $x=(x_t)_{t=1}^{N}$ and $u=(u_t)_{t=0}^{N-1}$. 
This problem is challenging due to the non-convexity of the cost function \eqref{eq:problem:cost}, the nonlinearity of the uncertain dynamics \eqref{eq:problem:dynamics}, the joint chance constraint \eqref{eqn:problem:joint_chance_constraint}, 
and the non-Markovianity of the state-trajectory  $(x_{t})_{t=1}^{N}$ which requires reasoning about the joint probability distribution of the state trajectory. 

Without imposing assumptions regarding the system parameterization and the distribution of the disturbances and parameters, the problem in \eqref{eqn: chance constrained optimization problem} is intractable.
Instead of relying on such assumptions, in this work, we assume access to a dataset of observed transitions from the system. 
Such a dataset may come from prior observations of the system evolution or from high-fidelity simulations. 

\begin{assumption}[Dataset]\label{assum:dataset}
    We have access to a dataset $\mathcal{D}=\lbrace (x_{0}^{i}, u^{i}, x^{i}) \rbrace_{i=1}^{M}$ of $M\in\mathbb{N}$ independent and identically distributed (i.i.d.) samples, 
    where  $x_{0}^{i}$ and $u^{i}$ are sampled i.i.d. from probability distributions on $\mathcal{X}$ and $\mathcal{U}^{N}$ 
    and $x^{i}=(x_1^i,{\dots},x_N^i)$ are i.i.d.\,trajectories satisfying \eqref{eqn: system dynamics} for each $i$.
\end{assumption}

In the next section, we propose an equivalent reformulation of the problem in \eqref{eqn: chance constrained optimization problem} using stochastic kernels. This reformulation is the basis for our data-driven approach that leverages the dataset $\mathcal{D}$.

\subsection{Reformulation using Stochastic Kernels}

The state trajectory $x$ that satisfies \eqref{eqn: system dynamics} can be characterized by a stochastic kernel $Q : \mathscr{B}(\mathcal{X}^{N}) \times \mathcal{X} \times \mathcal{U}^{N} \to [0, 1]$ that assigns a probability measure $Q(\cdot \mid x_{0}, u)$ to every initial condition $x_{0} \in \mathcal{X}$ and control sequence $u \in \mathcal{U}^{N}$ on the measurable space $(\mathcal{X}^{N}, \mathscr{B}(\mathcal{X}^{N}))$. 
This kernel can be defined as $Q(A \mid x_{0}, u) \triangleq \mathbb{P}(\bigwedge_{t=1}^{N} x_{t} =f_t(x_0,u,w,\theta)\in A_{t})$ for any $A = A_{1} \times \cdots \times A_{N}\in\mathscr{B}(\mathcal{X}^N)$ and  $(x_{0}, u) \in \mathcal{X} \times \mathcal{U}^{N}$. As such, Assumption \ref{assum:dataset} states that one has access to $M$ independent trajectory samples $x^i$ that are distributed according to $Q(\cdot \mid x_{0}^{i}, u^{i})$. 

For any initial state $x_{0} \in \mathcal{X}$, any control trajectory $u \in \mathcal{U}^{N}$, and any measurable function $g:\mathcal{X}^N\rightarrow\R$, we denote the expectation with respect to this probability measure as $\E_{Q(x|x_0,u)}[g(x)]=\int_{\mathcal{X}^N}g(x)Q(\textrm{d}x\mid x_0,u)$. 
Since $\E_{Q(x \mid x_0,u)}[g(x)]$ is a function of the initial condition $x_0$ and control input $u\in\mathcal{U}^N$, reformulating the original problem in \eqref{eqn: chance constrained optimization problem} amounts to computing the expectation over the stochastic control policy. Since the distribution of the control inputs $u$ is characterized by the stochastic kernel $\pi$, 
for any measurable function $h:\mathcal{U}^N\to\R$, we define the conditional expectation $\E_{\pi(u|x_0)}[h(u)]=\int_{\mathcal{U}^N}h(u)\pi(\textrm{d}u|x_0)$.  

The two stochastic kernels $Q$ and $\pi$ allow representing the joint chance constraint in \eqref{eqn:problem:joint_chance_constraint} as the expectation of an indicator function.
Let $\mathcal{T} = (\mathcal{X} \backslash \mathcal{O}_{1}) \times \cdots \times (\mathcal{X} \backslash \mathcal{O}_{N-1}) \times \mathcal{X}_{\textrm{goal}} \subset \mathcal{X}^{N}$ denote the set of all state trajectories which reach the goal set 
while avoiding the unsafe sets 
at all times, and let $\boldsymbol{1}_{\mathcal{T}}(x)$ denote the indicator of $\mathcal{T}$, which is one if $x_{N} \in \mathcal{X}_{\textrm{goal}}$ and $x_{t} \not\in \mathcal{O}_{t}$ for all $t = 1, \ldots, N-1$, and is zero otherwise. Then, the joint chance constraint in \eqref{eqn:problem:joint_chance_constraint} can be reformulated as 
\begin{equation}
    \label{eqn:problem:joint_chance_constraint expectation}
    \mathbb{E}_{\pi(u \mid x_{0})} \left[\mathbb{E}_{Q(x \mid x_{0}, u)}[\boldsymbol{1}_{\mathcal{T}}(x)]
    \right]
    = \mathbb{P} \biggl( \biggl( \bigwedge_{t=1}^{N-1} x_{t} \not\in \mathcal{O}_{t} \biggr) \land (x_{N} \in \mathcal{X}_{\textrm{goal}}) \biggr).
\end{equation}
Using \eqref{eqn:problem:joint_chance_constraint expectation} and the stochastic kernel $Q$,
we reformulate the original problem in \eqref{eqn: chance constrained optimization problem} as
\begin{subequations}
\label{eqn: equivalent chance constrained optimization problem}
\begin{align}
    \min_{\pi\in\Delta} \quad & \mathbb{E}_{\pi(u \mid x_{0})} \left[
    \mathbb{E}_{Q(x \mid x_{0}, u)} [\ell^{x}(x) + \ell^{u}(u)]
    \right]
    \\
    \textnormal{s.t.} \quad & \mathbb{E}_{\pi(u \mid x_{0})} \left[
    \mathbb{E}_{Q(x \mid x_{0}, u)}[\boldsymbol{1}_{\mathcal{T}}(x)]
    \right] \geq 1 - \delta.
\end{align}
\end{subequations}
Since the stochastic kernel $Q$ that characterizes the state trajectory $x$ is unknown, the problem above is generally intractable. 
In the next section, we show how the dataset $\mathcal{D}$ allows relaxing the problem in \eqref{eqn: equivalent chance constrained optimization problem} using a framework known as \emph{kernel embeddings of distributions}. This approach allows representing the expectation operators in \eqref{eqn: equivalent chance constrained optimization problem} as elements in a high-dimensional function space and subsequently reformulating the original problem as a linear program in this space.

\section{Hilbert Space Embeddings of Distributions}
\label{sec:stochastic_embeddings}

One of the primary challenges in solving \eqref{eqn: equivalent chance constrained optimization problem}
consists of evaluating the expectations with respect to the stochastic kernels $Q$ and $\pi$ which characterize dynamics uncertainty and the stochasticity of the control policy.  
Our key insight consists of embedding each integral operator as an element in a reproducing kernel Hilbert space (RKHS). Using this approach, evaluating the expectations in \eqref{eqn: equivalent chance constrained optimization problem} amounts to performing a linear operation between two functions of an RKHS. 
We leverage the dataset from Assumption \ref{assum:dataset} to approximate the expectation operator associated to the unknown stochastic kernel $Q$. Then, by selecting a finite-dimensional representation for the stochastic policy, we propose a tractable finite-dimensional approximation of the problem in \eqref{eqn: equivalent chance constrained optimization problem},

\subsection{Reproducing Kernel Hilbert Space Embeddings}

In this section, we endow our problem with an RKHS structure that enables representing the expectation operators in problem \eqref{eqn: equivalent chance constrained optimization problem} as linear functions. 
To do so, we first define the positive definite  kernel function $k_{\mathcal{X}^{N}} : \mathcal{X}^{N} \times \mathcal{X}^{N} \to \mathbb{R}$ \cite[Definition~4.15]{steinwart2008support}. 
According to the Moore-Aronszajn theorem \citep{aronszajn1950theory}, this kernel defines a unique corresponding RKHS $\mathscr{H}$ of functions from $\mathcal{X}^{N}$ to $\mathbb{R}$ with associated inner product $\langle \cdot, \cdot \rangle_{\mathscr{H}}$ such that: 
(i) $k_{\mathcal{X}^{N}}(x, \cdot) \in \mathscr{H}$ for all $x \in \mathcal{X}^{N}$, and (ii) $g(x) = \langle g, k_{\mathcal{X}^{N}}(x, \cdot) \rangle_{\mathscr{H}}$ for all $g \in \mathscr{H}$ and $x \in \mathcal{X}^{N}$.
As a consequence of the reproducing property, $k_{\mathcal{X}^{N}}(x, x^{\prime}) = \langle k_{\mathcal{X}^{N}}(x, \cdot), k_{\mathcal{X}^{N}}(x^{\prime}, \cdot) \rangle_{\mathscr{H}}$, which is often referred to  as the kernel trick \citep{steinwart2008support}.
We assume that the kernel $k_{\mathcal{X}^{N}}$ is $\mathscr{B}(\mathcal{X}^{N})$-measurable and bounded, such that $\sup_{x \in \mathcal{X}^{N}} \sqrt{k_{\mathcal{X}^{N}}(x, x)} < \infty$. 
With these conditions, according to \citep{song2009hilbert}, for every $x_0\in\mathcal{X}$ and $u\in\mathcal{U}^N$, there exists an element $m(x_{0}, u)\in\mathscr{H}$, called the \emph{conditional distribution embedding}, which is a linear function from $\mathcal{X}^N$ to $\R$ defined as
\begin{equation}\label{eq:conditional_distr_emb}
    m(x_{0}, u) \triangleq \mathbb{E}_{Q(x \mid x_{0}, u)}[k_{\mathcal{X}^{N}}(x, \cdot)].
\end{equation}
By the reproducing property, the expectation of any function of the RKHS $\mathscr{H}$ with respect to the stochastic kernel $Q$ can be evaluated as an inner product with the conditional distribution embedding. Specifically, for any $x_0\in\mathcal{X}$ and any $u\in\mathcal{U}^N$,
\begin{equation}\label{eq:conditional_expectation_Q_g}
\mathbb{E}_{Q(x \mid x_{0}, u)}[g(x)] = \langle g, m(x_{0}, u) \rangle_{\mathscr{H}} \quad \text{for any } g\in\mathscr{H}.
\end{equation}
Assuming that the cost function $\ell^{x}$ and the indicator function $\boldsymbol{1}_{\mathcal{T}}$ belong to $\mathscr{H}$, the inner expectations in \eqref{eqn: equivalent chance constrained optimization problem} with respect to $Q$ can be rewritten as a linear operation using the distribution embedding $m(x_{0}, u)$. 
As such, our approach does not rely on specific assumptions about the dynamics, the uncertain parameters, and the disturbances. The main challenge consists of computing an accurate approximation of the distribution embedding $m(x_0,u)$, which we pursue in the next section.

\subsection{Empirical Embedding Estimate}

Since the true stochastic kernel $Q$ is unknown a priori, we do not have access to the conditional distribution embedding $m(x_{0}, u)\in\mathscr{H}$. 
To construct an empirical estimate $\hat{m}(x_{0}, u)$ of this operator, we leverage the dataset $\mathcal{D}$ from Assumption \ref{assum:dataset}. 
Specifically, as in \citep{grunewalder2012conditional,caponnetto2007optimal,micchelli2005learning}, we search for a best-fit solution in the RKHS as the solution to the regularized least-squares problem
$\hat{m} = \arg \min_{f \in \mathscr{Q}} \frac{1}{M} \sum_{i=1}^{M} \lVert k_{\mathcal{X}^N}(x^{i}, \cdot) - f(x_{0}^{i}, u^{i}) \rVert_{\mathscr{H}}^{2} + \lambda \lVert \hat{m} \rVert_{\mathscr{Q}}^{2}$, where $\lambda > 0$ is a regularization parameter and $\mathscr{Q}$ is a vector-valued RKHS of functions from $\mathcal{X} \times \mathcal{U}^{N}$ to $\mathscr{H}$ \citep[see][for more information]{micchelli2005learning}. 
The solution to this problem is unique:
\begin{equation}
    \hat{m}(x_{0}, u) = \Phi^{\top} (G + \lambda M I)^{-1} K(x_{0}, u),
\end{equation}
where 
$\Phi \in \mathscr{H}^{M}$ is a feature vector with elements $\Phi_{i} = k_{\mathcal{X}^{N}}(x^{i}, \cdot)\in\mathscr{H}$, 
$G\in\mathbb{R}^{M\times M}$ is a Gram matrix with elements $G_{ij} = k_{\mathcal{X}}(x_{0}^{i}, x_{0}^{j}) k_{\mathcal{U}^{N}}(u^{i}, u^{j})$, where $k_{\mathcal{X}}$ and $k_{\mathcal{U}^{N}}$ are positive definite kernels on $\mathcal{X}$ and $\mathcal{U}^{N}$, respectively,
and 
$K(x_{0}, u) \in \mathbb{R}^{M}$ is a vector with elements $[K(x_{0}, u)]_{i} = k_{\mathcal{X}}(x_{0}^{i}, x_{0}) k_{\mathcal{U}^{N}}(u^{i}, u)$.
This estimate $\hat{m}(x_0,u)$ of the conditional distribution embedding $m(x_0,u)$ enables an efficient approximation of the expectation in \eqref{eq:conditional_expectation_Q_g}. Specifically, for any function $g\in\mathscr{H}$,
\begin{align}\label{eq:expectation_wrt_mhat}
\mathbb{E}_{Q(x \mid x_{0}, u)}[g(x)]
= \langle g, m(x_{0}, u) \rangle_{\mathscr{H}}
\approx
\langle g, \hat{m}(x_{0}, u) \rangle_{\mathscr{H}}
= \boldsymbol{g}^{\top} (G + \lambda M I)^{-1} K(x_{0}, u), 
\end{align} 
where $\boldsymbol{g} = [g(x^{1}), \ldots, g(x^{M})]^{\top}\in\mathbb{R}^M$. 
With \eqref{eq:expectation_wrt_mhat}, the inner expectations in \eqref{eqn: equivalent chance constrained optimization problem} with respect to the unknown stochastic kernel $Q$ can be approximated using simple matrix operations. 

The quality of this approximation generally depends on the dataset $\mathcal{D}$ from Assumption \ref{assum:dataset}, on the kernel choice, and on the number of samples $M$. 
The convergence properties of conditional distribution embeddings have been studied in the literature, see \citep{song2009hilbert, song2010nonparametric, grunewalder2012conditional, park2020measure}.  
Notwithstanding minor variations in the underlying theoretical framework, existing results show that the empirical estimate of a kernel distribution embedding converges in probability to the true embedding as the sample size increases, and suggest an optimal rate of convergence of $\mathcal{O}_{p}(M^{-1/4})$ \citep{park2020measure}.
Exactly quantifying the approximation error in \eqref{eq:expectation_wrt_mhat} for a finite number of samples is beyond the scope of this work.

\subsection{Stochastic Control Trajectory Representation}

\begin{figure}[t]
\centering
\includegraphics[width=0.5\linewidth,trim=0 0 0 0, clip]{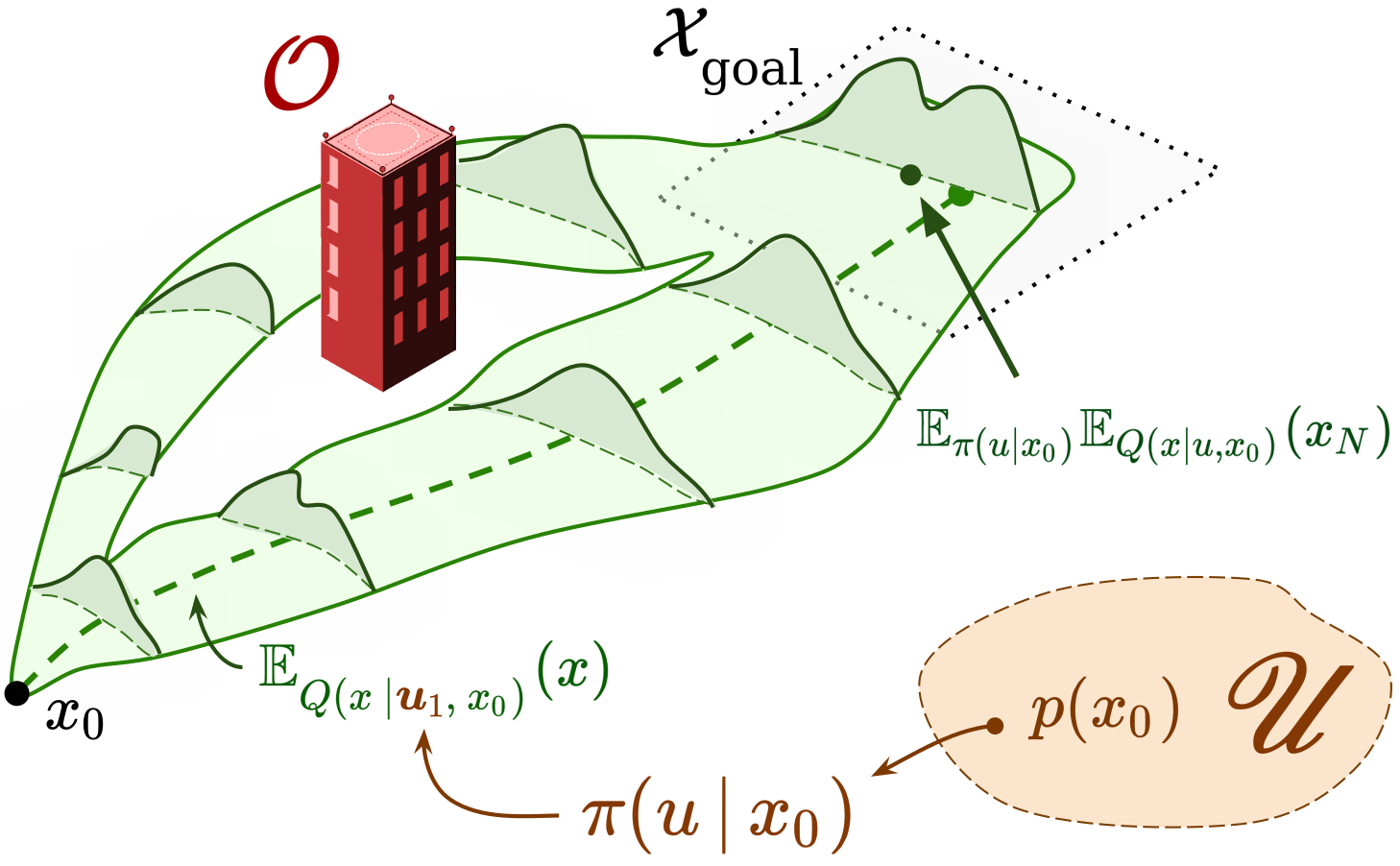}
	\caption{We search for a stochastic kernel embedding \textcolor[rgb]{0.43, 0.21, 0.1}{$p(x_0)$} in the RKHS \textcolor[rgb]{0.43, 0.21, 0.1}{$\mathscr{U}$}. 
	This defines a stochastic kernel  \textcolor[rgb]{0.43, 0.21, 0.1}{$\pi(u|x_0)$} that characterizes the control policy. Using this policy kernel and the approximate state trajectory stochastic kernel \textcolor[rgb]{0,0.4,0}{$\hat{Q}(x|x_0,u)$}, we represent the state trajectory distribution and reformulate the original stochastic control problem. }
	\vspace{-5mm}
\end{figure}
	
To reformulate the problem in \eqref{eqn: equivalent chance constrained optimization problem}, it remains to compute the expectation with respect to any chosen stochastic policy $\pi$. 
At this point, one could search for a deterministic control sequence $u\in\mathcal{U}^N$ and obtain a tractable approximate reformulation of the original problem. However, this reformulation would be non-convex if $\ell^u$ and $k_{\mathcal{U}^{N}}$ 
are non-convex.

Instead, optimizing over the larger set of stochastic policies allows for more efficient control strategies \citep{Altman1999,Masahiro2016}. 
Note that under the condition that $k_{\mathcal{U}^{N}}$ is $\mathscr{B}(\mathcal{U}^{N})$-measurable and bounded, any stochastic policy $\pi(\cdot \mid x_0)$ can be represented by a conditional distribution embedding $p(x_0)$ in the RKHS $\mathscr{U}$ associated with $k_{\mathcal{U}^{N}}$. 
Accordingly, the expectation of any function $h\in\mathscr{U}$ with respect to the stochastic policy can be expressed as an inner product in the RKHS $\mathscr{U}$ as 
$\mathbb{E}_{\pi(u \mid x_{0})}[h(x_0,u)] = \langle h, p(x_{0}) \rangle_{\mathscr{U}}$. 
However, depending on the choice of kernel $k_{\mathcal{U}^{N}}$, the representation of the embedding $p(x_0)$ may be infinite-dimensional. 
Thus, optimizing over all possible stochastic embeddings is generally intractable. 
This observation motivates our finite-dimensional policy representation, which we characterize by the set of stochastic kernel embeddings
\begin{equation}
    \label{eqn: policy approximation}
    p(x_{0}) = \sum_{j=1}^{P} \gamma_{j}(x_{0}) k_{\mathcal{U}^{N}}(\tilde{u}^{j}, \cdot),
    \ \ 
    \text{where } \  
    \sum_{j=1}^{P} \gamma_{j}(x_{0}) = 1,
    \ \  
    0 \preceq \gamma(x_{0}),
\end{equation}
where $\gamma(x_{0})\in\R^P$, 
the elements $\tilde{u}^j\in\mathcal{U}^N$ are user-specified control sequences, and 
$j=1,\dots,P$. 
The finite set of control sequences $\mathcal{A} = \smash{\lbrace \tilde{u}^{j} \rbrace_{j=1}^{P}}$ can be chosen strategically to uniformly cover the control space $\mathcal{U}^N$, can be sampled randomly (independently of Assumption \ref{assum:dataset}), or can be pre-specified depending on the application (e.g., informed by a sampling-based planner with an approximate deterministic dynamics model, \citealp{IchterHarrisonEtAl2018}). 
The coefficients $\gamma_j(x_{0})$ characterize the probability values that weight the admissible control sequences in $\mathcal{A}$. 
Correspondingly, the last two conditions in \eqref{eqn: policy approximation} constrain the coefficient vector $\gamma(x_{0})$ to lie within a probability simplex $\mathscr{S} = \lbrace \gamma\in\R^P \mid \sum_{i=1}^P \gamma_i = 1, 0 \preceq \gamma \rbrace$. 
This guarantees that the kernel distribution embedding $p(x_0)$ defines a valid stochastic kernel $\pi$ characterizing the control policy.

Restricting the search of stochastic kernel embeddings to a finite-dimensional subspace of $\mathscr{U}$ allows a tractable relaxation of the original problem. Indeed,  for any function $h\in\mathscr{U}$, we have
\begin{equation}\label{eq:expectation_wrt_pihat}
    \mathbb{E}_{\pi(u \mid x_{0})}[h(u)]  = 
    \left\langle h, p(x_{0}) \right\rangle_{\mathscr{U}}
    =
    \left\langle h, \sum_{j=1}^P\gamma_j(x_0)k_{\mathcal{U}^{N}}(\tilde{u}^{j}, \cdot) \right\rangle_{\mathscr{U}}
    =
    \sum_{j=1}^P h(\tilde{u}^j)\gamma_j(x_0).
\end{equation}
Assuming that the cost function $\ell^u\in\mathscr{U}$, this relaxation enables a tractable evaluation of the expectation operator with respect to the stochastic control policy $\pi(\cdot\mid x_0)$. 
Intuitively, $\pi(\cdot\mid x_0)$ is a sum of Dirac functions centered at each control trajectory $\tilde{u}^j\in\mathcal{U}^N$ weighted by the probability coefficient $\gamma_j(x_0)$ \citep{berlinet2011reproducing}.  
Importantly, \eqref{eq:expectation_wrt_pihat} is linear in the unknown parameter $\gamma(x_0)\in\R^P$. 
This relaxation offers a tractable linear reformulation of the original problem, although it induces sub-optimality relative to the computationally intractable problem of optimizing over the infinite-dimensional space of stochastic policies. We leave the quantification of this sub-optimality gap (perhaps by first quantifying the error between the empirical approximation in \eqref{eqn: policy approximation} and any feasible stochastic embedding $p(x_0)$, e.g., see \citep[Section 6.2]{kanagawa2018gaussian}) to future work.

\subsection{Approximate chance constrained Optimization Problem}

Using the estimate $\hat{m}(x_{0}, u)$ and the policy estimate $p(x_{0})$, the expectations in \eqref{eqn: equivalent chance constrained optimization problem} can be approximated using simple matrix multiplications and inner products. Specifically, 
for any function $g\in\mathscr{H}$, 
combining \eqref{eq:expectation_wrt_mhat} with \eqref{eq:expectation_wrt_pihat}, we obtain 
\begin{align}
    \mathbb{E}_{\pi(u\mid x_0)}[
    \mathbb{E}_{Q(x\mid x_0,u)}[
        g(x)
    ]]
    \approx
    \langle \langle g, \hat{m}(x_{0}, \cdot) \rangle_{\mathscr{H}}, \hat{p}(x_{0}) \rangle_{\mathscr{U}}
    =
    \sum_{j=1}^P 
    \boldsymbol{g}^{\top} (G + \lambda M I)^{-1} K(x_{0},\tilde{u}^j)
    \gamma_j(x_0).
\end{align}
\\[-6mm]
This motivates the following approximation of the problem in \eqref{eqn: equivalent chance constrained optimization problem}: 
\begin{subequations}
\label{eqn: approximate chance constrained optimization problem}
\begin{align}
    \min_{\gamma(x_{0}) \in \mathbb{R}^{P}} \quad & 
    \left(\boldsymbol{\ell}^{x}{}^{\top} (G + \lambda M I)^{-1} R(x_{0})
    + \boldsymbol{\ell}^{u}{}^{\top} \right)\gamma(x_{0}) \\
    \label{eqn: approximate chance constraint}
    \textnormal{s.t.} \quad & \left(\bar{\boldsymbol{1}}_{\mathcal{T}}^{\top} (G + \lambda M I)^{-1} R(x_{0}) \right) \gamma(x_{0}) \geq 1 - \delta, \\
    & \boldsymbol{1}^{\top} \gamma(x_{0}) = 1, \quad
    0 \preceq \gamma(x_{0}),
\end{align}
\end{subequations}
where $R(x_{0}) \in \mathbb{R}^{M \times P}$ is a matrix such that each column $[R(x_{0})]_{j} = K(x_{0}, \tilde{u}^{j})$ such that $[R(x_{0})]_{ij} = k_{\mathcal{X}}(x_{0}^{i}, x_{0}) k_{\mathcal{U}^{N}}(u^{i}, \tilde{u}^{j})$, 
and 
$\boldsymbol{\ell}^{x},\boldsymbol{\ell}^{u}, \bar{\boldsymbol{1}}_{\mathcal{T}}\in\mathbb{R}^M$ are vectors with elements $\boldsymbol{\ell}_{i}^{x} = \ell^{x}(x^{i})$, $\boldsymbol{\ell}_{i}^{u} = \ell^{u}(\tilde{u}^{j})$, and $[\bar{\boldsymbol{1}}_{\mathcal{T}}]_{i} = \boldsymbol{1}_{\mathcal{T}}(x^{i})$, respectively. 
The approximate problem in \eqref{eqn: approximate chance constrained optimization problem} is a linear program over the coefficients $\gamma(x_{0})$ (that represent probability weights over the admissible control sequences in $\mathcal{A}$) that can be efficiently solved via interior-point or simplex algorithms. 
Thus, the solution to \eqref{eqn: approximate chance constrained optimization problem} is a randomized control strategy, i.e., an open-loop stochastic control policy with values in $\mathcal{A}$. 
The quality of the resulting stochastic control strategy that solves the problem in \eqref{eqn: approximate chance constrained optimization problem}  generally depends on the quality of the dataset $\mathcal{D}$ from Assumption \ref{assum:dataset} and on the choice of the control inputs in $\mathcal{A}$.

\section{Numerical Experiments}
\label{section: numerical results}

\begin{figure}[t]
    \centering
    \vspace{-4mm}
    \includegraphics[keepaspectratio, height=1.6in,trim=0 0 0 0, clip]{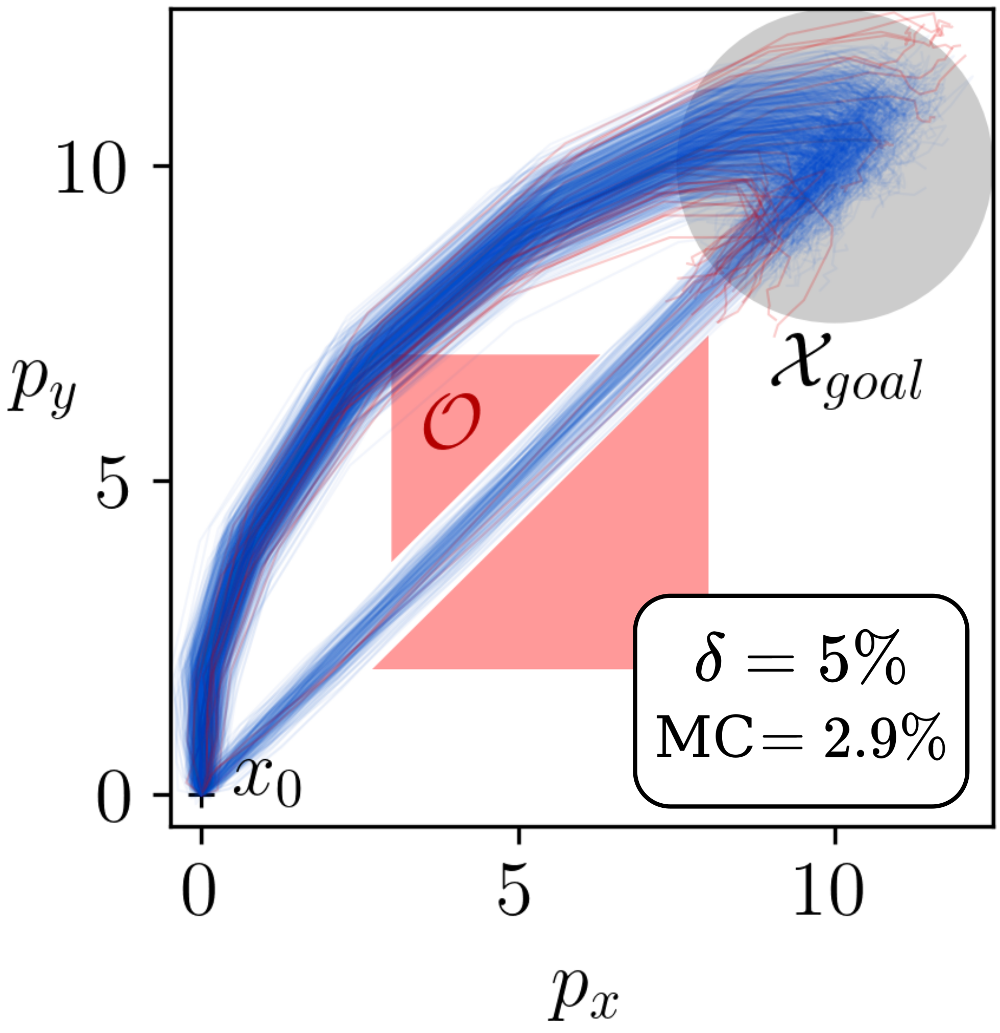}
    \hspace{2mm}
    \includegraphics[keepaspectratio, height=1.6in,trim=0 0 0 0, clip]{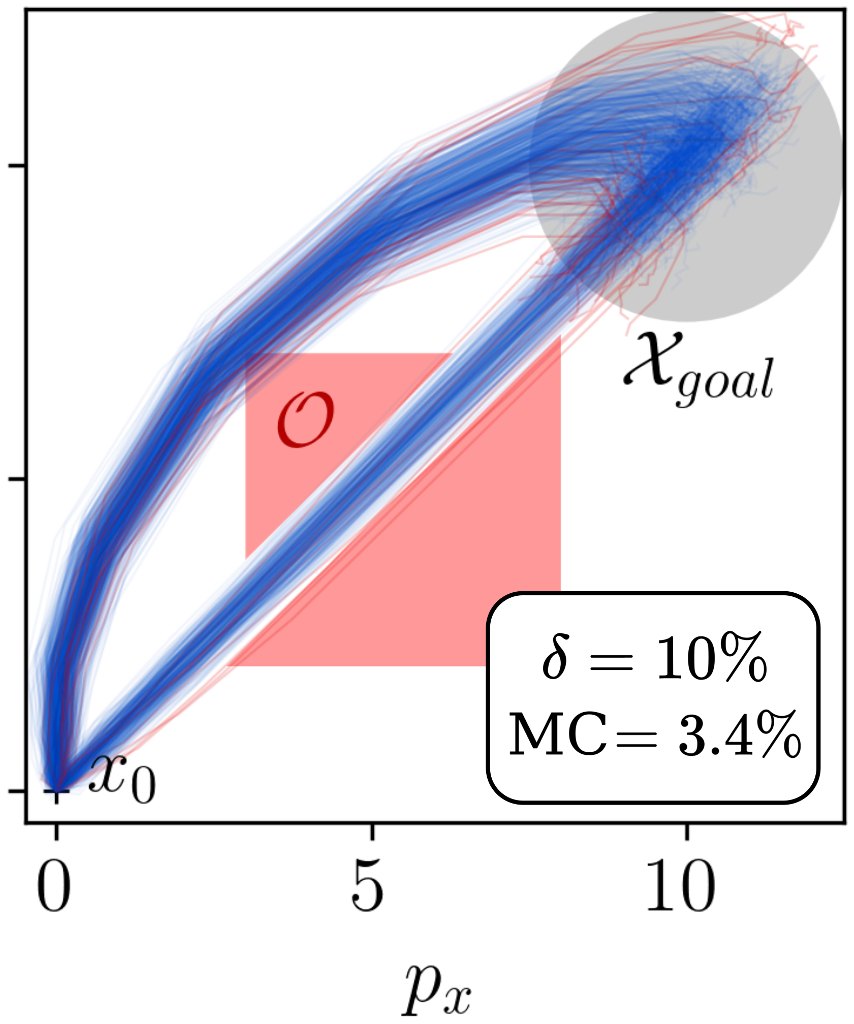}
    \hspace{2mm}
    \includegraphics[keepaspectratio, height=1.6in,trim=0 0 0 0, clip]{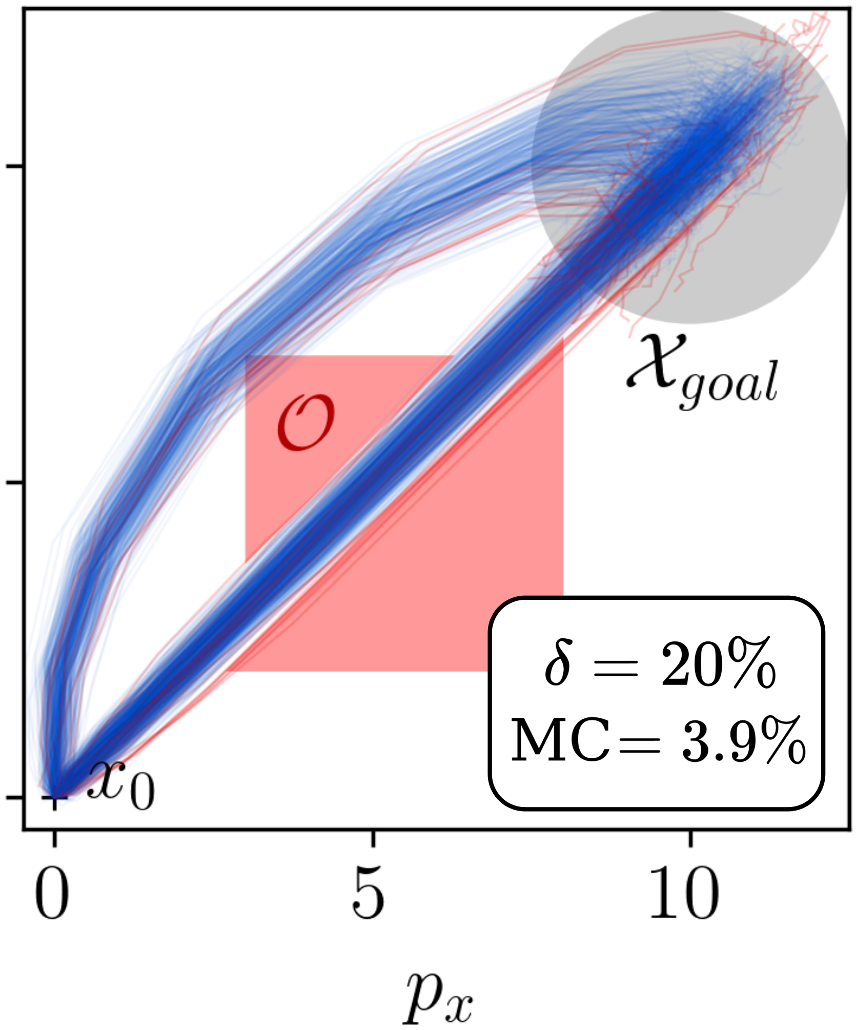}
    \includegraphics[keepaspectratio, height=1.6in,trim=0 0 0 0, clip]{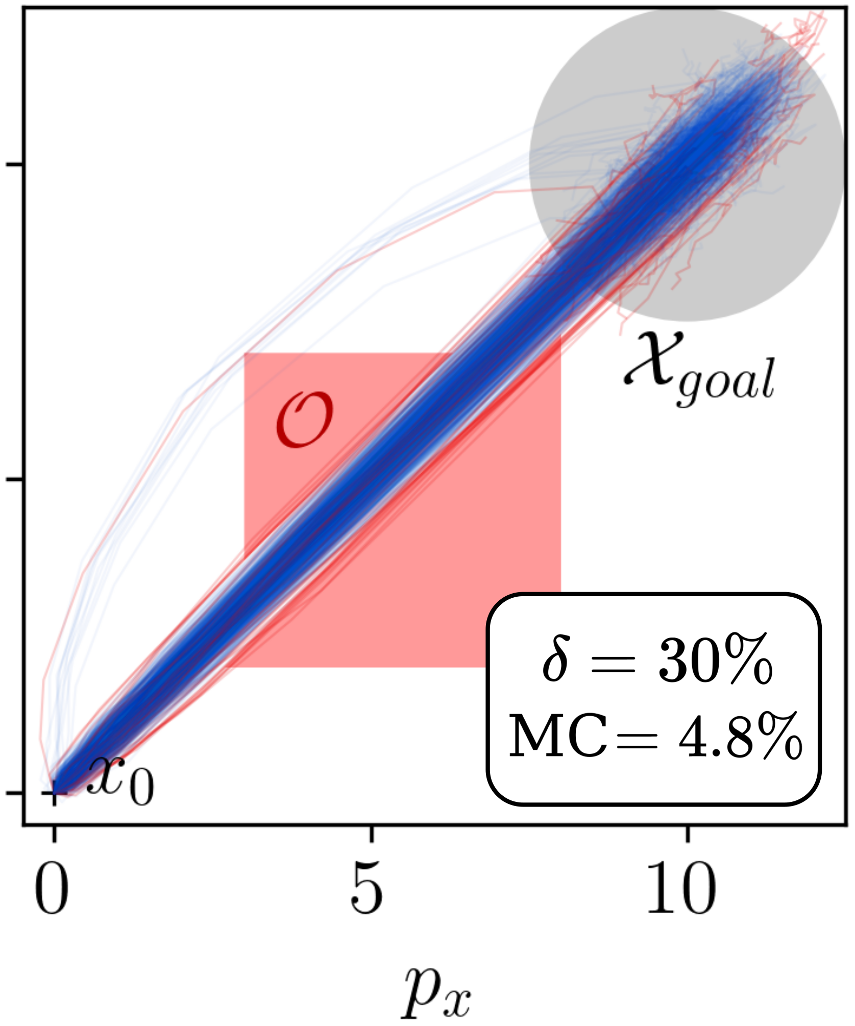}
    \vspace{-2mm}
    \caption{Solutions from our proposed data-driven algorithm for different values of the tolerable failure probability threshold $\delta$. Blue trajectories from Monte-Carlo (MC) simulations denote feasible trajectories which reach the goal set $\mathcal{X}_{\textrm{goal}}$ and avoid obstacles $\mathcal{O}$. Red trajectories violate constraints.} 
    \label{fig:results:kernel:deltas}
    \vspace{-4mm}
\end{figure}

We consider the problem of controlling a quadrotor system operating in turbulent conditions, with nonlinear dynamics given by $x_{t+1} = A
    x_{t} + 
    B(\theta)
    u_{t} + 
    d(x_t, \theta) + 
    w_{t}$, where
\begin{align}
    \label{eqn: repeated integrator dynamics}
    A = 
    \scalebox{0.85}{$\begin{bmatrix}
        1 & \Delta t & 0 & 0 \\
        0 & 1 & 0 & 0 \\
        0 & 0 & 1 & \Delta t \\
        0 & 0 & 0 & 1
    \end{bmatrix}$},
    \ 
    B(\theta) = 
    \frac{1}{m}
    \scalebox{0.85}{$\begin{bmatrix}
        \Delta t^2/2 & 0 \\
        \Delta t & 0 \\
        0 & \Delta t^2/2 \\
        0 & \Delta t
    \end{bmatrix}$},
    \ 
    d(x_{t}, \theta) = -\alpha 
    \scalebox{0.85}{$\begin{bmatrix}
        \Delta t^2|v_x|v_x/2\\
        \Delta t|v_x|v_x\\
        \Delta t^2|v_y|v_y/2\\
        \Delta t|v_y|v_y
    \end{bmatrix}$},
\end{align}
and $\Delta t = 0.1$. 
The state and control inputs of the system are denoted as $x_{t} = [p_{x,t}, v_{x,t}, p_{y,t}, v_{y,t}] \in \mathbb{R}^{4}$ and $u_t = [u_x, u_y] \in \mathbb{R}^{2}$. 
Starting from an initial condition $x_{0} = [-0.5, 0, -0.5, 0]$, the goal consists of computing a stochastic policy such that the system reaches the set $\mathcal{X}_{\textrm{goal}} = \lbrace x \in \mathbb{R}^{4} \mid \|(p_x{-}10,p_y{-}10)\| \leq 2.5$ at time $N=15$ while avoiding two polytopic obstacles $\mathcal{O}$ shown in Figure \ref{fig:results:kernel:deltas}. 
The dynamics are parameterized by uncertain parameters $\theta=(m,\alpha)$ representing the mass $m>0$ of the system and an uncertain drag coefficient $\alpha>0$.
We assume that the parameters $(m,\alpha)$ of the system are uncorrelated random variables such that $(m-0.75)/(0.5)\sim\textrm{Beta}(2,2)$ and 
$(\alpha-0.4)/(0.2)\sim\textrm{Beta}(2,5)$, where $\textrm{Beta}(a,b)$ denotes a Beta distribution with shape parameters $(a,b)$. 
As discussed in Section \ref{section: problem formulation}, this system is not Markovian due to the temporal correlation between the state trajectory
$x$ and the uncertain parameter $\theta$. 

We generate a dataset $\mathcal{D} = \lbrace (x_{0}^{i}, u^{i}, x^{i}) \rbrace_{i=1}^{M}$ as in Assumption \ref{assum:dataset} with $M=2500$ by uniformly sampling initial conditions $x_0^i \in [-0.5, -0.05, -0.5, -0.05] \times [0.5, 0.05, 0.5, 0.05]$. 
The control sequences $u^i$ are generated by uniformly sampling $u^i_t\in[0,1]^2$ for $t=0,1,2$ to impose variability in the sample trajectories, and then selecting controls for $t\geq 3$ via a feedback controller, where $u^i_t=K(x_t-(10,0,10,0))$ given closed-loop trajectories $x$ simulated using \eqref{eqn: repeated integrator dynamics} and a pre-specified linear feedback gain $K\in\R^{2\times 4}$. 
These control trajectories $u^i$ are then used to simulate open-loop trajectories $x^i$ again, which ensures that these trajectories correspond to open-loop control inputs. 
Finally, the control sequences \smash{$\mathcal{A} = \lbrace \tilde{u}^{j} \rbrace_{j=1}^{P}$} with $P=2500$ are selected with the same approach as described above, except that we use a uniform grid to select the control inputs $\tilde{u}^j_t$ at the first three time steps, $t = 0, 1, 2$, and use the dynamics in \eqref{eqn: repeated integrator dynamics} with constant parameters $(m, \alpha) = (1, 0.005)$. 
We then presumed no knowledge of the system dynamics in \eqref{eqn: repeated integrator dynamics} or the stochastic kernel $Q$ to compute a solution to the problem.

Using the datasets $\mathcal{D}$ and $\mathcal{A}$, we compute a kernel-based approximation of the original chance constrained problem and stochastic policy using \eqref{eqn: approximate chance constrained optimization problem}. 
We set the regularization parameter to be $\lambda=10^{-7}$ and use a Gaussian kernel $k(x, x^{\prime}) = \exp(-\sigma \lVert x - x^{\prime} \rVert_{2}^{2})$ with the kernel parameter $\sigma$ chosen as the median Euclidean distance between sample points.  
We use the \texttt{linprog} function from \texttt{scipy} to formulate and solve the linear programs in \eqref{eqn: approximate chance constrained optimization problem}.
Our Python implementation, as well as code to reproduce all figures and analysis is available at \scalebox{0.9}{\url{https://github.com/ajthor/socks}}.

Results are shown in Figure \ref{fig:results:kernel:deltas} for different values of $\delta$. Figure \ref{fig:results:kernel:deltas} shows $1{,}000$ Monte-Carlo simulations of our system using the open-loop policy computed using our method.
For all values of $\delta$, the algorithm computes a randomized control policy that satisfies the desired success rate $1-\delta$. 
The results in Figure \ref{fig:results:kernel:deltas} show that with larger values of $\delta$, the algorithm is more likely to select control inputs that generate trajectories passing through the riskier middle corridor between the obstacles. 
This is the expected behavior as $\delta$ directly controls the maximum allowable probability of constraint violation. This shows that our approach is able to compute a mixed policy that solves the problem. 

\begin{wrapfigure}{!R}{0.4\linewidth}
	\begin{minipage}{0.99\linewidth}
	\vspace{-4mm}
	\centering
\includegraphics[width=0.7\linewidth,trim=0 0 0 0, clip]{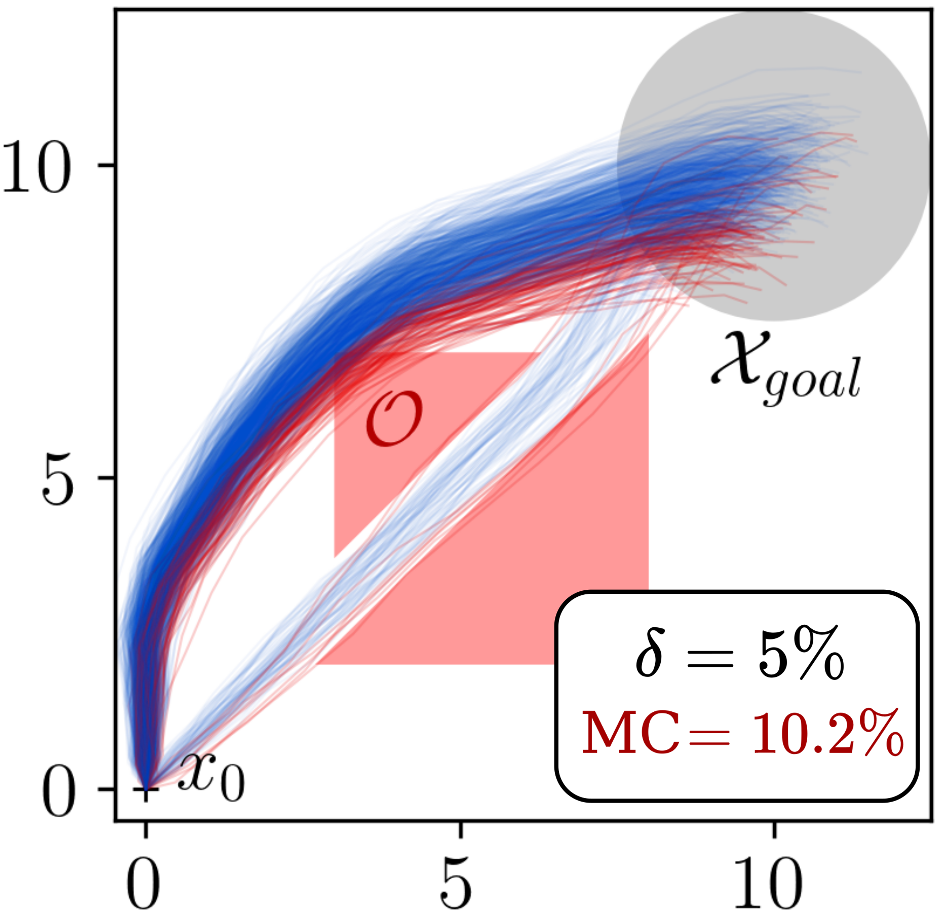}
	\vspace{-2mm}
	\caption{Solution from the approach proposed in \citep{Masahiro2016}. Blue Monte-Carlo trajectories denote feasible trajectories that reach the goal set $\mathcal{X}_{\textrm{goal}}$ and avoid obstacles $\mathcal{O}$. Red trajectories indicate trajectories which violate constraints. }
\label{fig:results:comparison_with_Ono}
	\vspace{-6mm}
	\end{minipage}%
	\end{wrapfigure}
We compare our approach with the method proposed in \citep{Masahiro2016}, which consists of decomposing the joint chance constraint using Boole's inequality and combining a Lagrangian relaxation approach with a bisection technique to obtain a mixed control strategy. 
Since this approach considers linear dynamics, we implement this baseline assuming that $m=1$ and $\alpha=0$ using the dynamics in \eqref{eqn: repeated integrator dynamics}. 
We present results for $\delta=0.05$ in Figure \ref{fig:results:comparison_with_Ono} and generate $1{,}000$ Monte-Carlo simulations using the resulting mixed control strategy and the nonlinear dynamics in \eqref{eqn: repeated integrator dynamics}. 
We observe that the solution from this approach is similar to the output of our data-driven method, in that it mixes two control strategies that generate trajectories from two different control sequences. 
However, 10.2\% of trajectories violate constraints using the method in \citep{Masahiro2016}. 
This comparison demonstrates the need to consider the nonlinear, non-Markovian dynamics of this system to reliably solve this problem.

\section{Conclusion}
\label{section: conclusion}

In this paper, we presented a novel data-driven technique leveraging stochastic kernel embeddings to compute stochastic control trajectories for general joint chance constrained control problems. 
By directly approximating the integral operator associated to the distribution of the state trajectory, this model-free approach is agnostic to the particular parameterization of the dynamics and of the distribution of the uncertain parameters and external disturbances. It is also computationally efficient, since the original stochastic problem is approximately reformulated as a linear program over the control parameters. Numerical experiments demonstrate the efficacy of our approach for controlling a challenging nonlinear non-Markovian dynamical system subject to non-convex constraints.

An important direction of future work is the accurate characterization of the approximation error in \eqref{eq:expectation_wrt_mhat}. Specifically, given further distributional and smoothness assumptions, finite-sample error bounds between the true and empirical embeddings in \eqref{eq:expectation_wrt_mhat} could allow the strict satisfaction of the original joint chance constraint \eqref{eqn:problem:joint_chance_constraint}. 
Future work will investigate different control input selection strategies, e.g., using non-convex optimization or cross-entropy policy optimization approaches. 
Of practical interest is investigating different choices of kernels to yield better accuracy and less conservative behavior (e.g., by meta-learning a kernel tailored to the true system).
Finally, our data-driven approach could motivate reinforcement learning strategies to actively gather data and improve the stochastic kernel embedding approximation quality while balancing the exploration-exploitation trade-off that arises from this constrained Markov decision process formulation.

\begin{acks}
The authors thank Masahiro Ono for insightful discussions about using mixed strategies for joint chance-constrained control and Apoorva Sharma for his helpful feedback. 
    This material is based upon work supported by the National Science Foundation under NSF Grant Numbers CNS-1836900 and CMMI-271742.  Any opinions, findings, and conclusions or recommendations expressed in this material are those of the authors and do not necessarily reflect the views of the National Science Foundation.
    The NASA University Leadership initiative (Grant \#80NSSC20M0163) provided funds to assist the authors with their research, but this article solely reflects the opinions and conclusions of its authors and not any NASA entity.
    This research was supported in part by the Laboratory Directed Research and Development program at Sandia National Laboratories, a multimission laboratory managed and operated by National Technology and Engineering Solutions of Sandia, LLC., a wholly owned subsidiary of Honeywell International, Inc., for the U.S. Department of Energy’s National Nuclear Security Administration under contract DE-NA-0003525.  The views expressed in this article do not necessarily represent the views of the U.S. Department of Energy or the United States Government.
\end{acks}

\bibliography{ASL_papers.bib,bib_2.bib,bibliography.bib}

\end{document}